# THE STRUCTURE AND CAGE FILLING OF GAS HYDRATES AS ESTABLISHED BY SYNCHROTRON POWDER DIFFRACTION DATA

Christiane D. Hartmann, Susanne Hemes, Andrzej Falenty and Werner F.Kuhs[*]
Geowissenschaftliches Zentrum der Universität Göttingen
Abteilung Kristallographie,
Goldschmidtstraße 1, 37077 Göttingen
GERMANY

## ABSTRACT

Synchrotron powder diffraction data were measured on various gas hydrates prepared at different fugacities and recovered to low temperatures. The data extend to sin $\theta/\lambda$ values of at least 1.01 Å$^{-1}$, in some cases up to 1.48 Å$^{-1}$, i.e. at least 3 times further in reciprocal space than the best single crystal data obtained so far. Here the results of $CO_2$-hydrate are presented in detail. Structural information is obtained from Rietveld refinements to unprecedented precision. All H-bonded O-O distances are found to be close to 2.75-2.76 Å, i.e. very close to the values in ice Ih; differences between a hydrogenated and deuterated host lattice are insignificant. Cage occupancies were obtained for the first time in free refinements together with positional parameters of the guest molecule and their atomic displacement parameters. There is good agreement between the experimental cage occupancies and prediction results from CSMGem.

## INTRODUCTION

Gas hydrates are non-stoichiometric compounds of hydrogen-bonded water molecules organized in form of cages around guest molecules that stabilize the lattice [1]. The most common cubic structures sI and sII are composed of two types of cavities: large (LC) and small (SC). For a full understanding of the physical chemistry of these compounds the cage filling is one of the most important structural parameters. There is multiple evidence that the cages are not necessarily fully occupied, yet a precise determination of cage occupancies has proven to be quite difficult. Spectroscopic methods in general do not provide access to the absolute cage fillings and crystallographic methods are bound with problems due to the extensive disorder of both host lattice and guest position. Particularly severe are the parameter correlations between guest positional coordinates, guest position occupancy and the thermal displacement parameters of the guests. In order to disentangle these correlations usually one needs to fix some of these parameters which in turn will cast doubts on the remaining freely refined values. We will show further below that some of these correlations can be overcome by using very high quality synchrotron powder diffraction data extending to large scattering angles. In the following we shall look in particular at the often investigated case of $CO_2$ hydrate due to its complex disorder situation but will also mention briefly some other hydrates of importance.

The first structural study of (deuterated) $CO_2$ hydrates quoting cage occupancies was published in 1998 based on *in situ* neutron powder diffraction data [2]. While there was general agreement that the large cage should be almost completely filled, the small cage situation was unclear at that time. The structural model with its positional parameters and the displacement parameters had to be fixed in this preliminary study. A guidance for the structural model of the guest atoms in the large cage was obtained from NMR work [3], while *ad hoc* assumptions had to be made for the SC, disordering the oxygen atoms along the [100] direction. With these input parameters the LC was found to be completely filled within error and the SC partly filled with values between 0.55 and 0.70, depending on the formation conditions and on the data set. The pressure (fugacity) dependence of the filling was also studied in this work. The small cage occupancy was found to be in disagreement with the predictions from the Langmuir models of Munck et al. [4] and Parrish and Prausnitz [5], deviating towards smaller values.

[*] Corresponding author, phone +49 551 393891, fax: +49 551 393891, e-mail: wkuhs1@gwdg.de

In the following years several crystallographic studies were performed starting with Ikeda et al. [6, 7], using neutron powder diffraction on deuterated samples at ambient pressure conditions up to temperatures slightly higher than 200K. The $CO_2$ model for the LC was again taken from [3] while for the SC four different models were tested, all with the carbon atom placed at the cage center. The model with the $CO_2$ oxygen atoms pointing to the water oxygens of the cage was found to give the best agreement with the diffraction data. Unfortunately, less attention was paid to the cage occupancies; values of 1.0 and 0.99 for LC and SC respectively are quoted without errors given, usually indicating that the values were not freely refined. The cage filling was explicitly considered in the work by Henning et al. [8], again on a deuterated sample using neutron diffraction. Supported by trial fits they assume a disorder of $CO_2$ molecules in the SC over 6 different orientations and for the LC a model disordered over 4 sites, as deduced from difference Fourier maps. The positional parameters of this model were freely refined and resulted in very reasonable C-O distances of 1.18(3) Å for the LC and 1.166(9) Å for the SC. Attempts to refine freely the cage occupancies failed due to correlations with the atomic displacement parameters and only approximate values were given. At the same time Udachin et al. [9] reported single crystal structure refinements of deuterated $CO_2$ hydrates based on X-ray diffraction data. The model was based on NMR work of the authors different from the earlier model [3], but more appropriate at the lower temperatures of the data collection. Free refinements of both, cage occupancies and displacement parameters were not possible for the SC and the acceptable range of the occupancy factor was estimated by varying the displacement parameters between reasonable limits to be centered at 0.71(2). A further neutron diffraction study of Circone et al. [10] did not allow for a determination of the cage occupancies. More recently Takeya et al. [11], using powder X-ray diffraction, determined the cage occupancies of LC and SC as 0.99 and 0.69 respectively, with an estimated error of 0.02. A direct space method was used in this work which varies trial structures (with fixed host lattice and varying guest models) in an attempt to match the experimental powder pattern. The method proved to be robust and was found to drive into the same solution starting from different trial settings. This method delivers model-independent results in such disordered systems and can provide reliable information on cage-fillings. Nevertheless, straight site disorder models obtained from free refinements are still not obsolete, as they are needed for the batch treatment of time-resolved experiments [12] in which thousands of data sets must be handled sequentially in a computationally economic way via disorder models. Thus we attempt in the following to tackle the "disorder problem" at a different point by overcoming some of the data limitations of neutron diffraction or laboratory X-ray sources. Using synchrotron powder diffraction up to very high scattering angles we can show that disorder models can be handled without serious correlation problems. Synchrotron powder diffraction data provide excellent signal-to-noise ratio at high scattering angles, combined with the unaffectedness of the strong low-angle Bragg reflections by extinction effects and are shown in the following to be superior to laboratory single crystal data. Low-angle reflections have a strong leverage on the cage occupancies but need reliable information at high scattering angles to develop their full potential. At high angles leverage for atomic displacement parameters is highest, in a specific range, depending on the mean-square-displacement under consideration [13]. Reliable information on displacement parameters permit to break parameter correlations between occupancies and displacements, often mediated via the scale factor. Furthermore, we also compare the various existing models in their ability to fit our synchrotron powder diffraction data. Finally, since we have produced samples at different fugacities, we are also able to compare the experimental degree of cage fillings with expectations from statistical thermodynamic theory.

# EXPERIMENTAL METHODS
## Sample preparation and data collection

The majority of clathrate samples were prepared starting from $H_2O$ or $D_2O$ (99.9% deuterated) ice Ih, which was obtained by spraying water into liquid nitrogen resulting in ice spheres with a typical diameter of several tens of µm [14]. In a few cases $D_2O$ frost was used for the formation of $CO_2$-hydrates at low temperatures (180 – 195 K) and pressures. This particular material was obtained through a controlled deposition of $D_2O$ water vapor on a cold rotating copper-plate partially submerged in the liquid nitrogen. Condensed frost was scraped off in regular intervals into liquid $N_2$ [15]. The production of deuterated ice took place in a glove box under dry nitrogen atmosphere to avoid isotopic exchange of the deuterated ice with atmospheric $H_2O$ [15]. The ice samples were placed into Al-cans with an inner diameter of 6.7 mm or in larger PFA-jars and inserted into pre-cooled aluminum pressure cells. The loaded cell was screwed onto the cryostat stick with its high pressure capillary linked to the gas-handling system [16]. Immediately afterwards the cell was immersed into a cold bath and left there for about 10 min. for equilibration at the desired formation temperature (Table I). Subsequently, the gas pressure (Purity of 99.998% $CO_2$) was raised in a few seconds to the value of the chosen formation pressure (Table 1). The pressure was monitored using Ashcroft pressure gauges calibrated with a mechanical high-precision Heise manometer. An initial rapid pressure drop marked the onset of clathrate formation. To ensure a formation close to the target pressure the gas was readjusted manually over the whole reaction period of a few weeks (Table I; the typical range of readjustments was 1-2 bars for the hydrates synthesized at high temperatures). At the end of the formation process the cell was cooled down in liquid $N_2$ until the rapid pressure fall in the reaction cell was registered. In the following 30 – 40 s the remaining gas was released and cell opened. The samples were immediately quenched in liquid nitrogen and subsequently stored in a Mover Dewar.

The samples for the synchrotron powder diffraction experiments were gently grinded under liquid $N_2$ and transferred into small quartz glass capillaries with an inner diameter of 1.0 – 1.7 mm. The powder diffraction data were acquired on the high-resolution diffractometer ID31 at ESRF (Grenoble) equipped with a nine crystal multi-analyzer stage [17]. The wavelength was determined to be 0.403027 Å by using a silicon standard powder. For data collection the samples were mounted vertically to the synchrotron beam (Bragg-Brentano geometry; 2θ-range of at least 0-48° 2θ, in some cases 0-100° 2θ) and spun with 300 rpm for good grain statistics. In addition measurements from different sample positions and within different 2θ-ranges were taken and merged together. To avoid sample decomposing and to achieve a low-noise diffraction pattern the samples were cooled by a coaxial $N_2$ stream set to a nominal temperature of 100 K (Figure 1a). This temperature was actually not achieved at the sample as deduced from the refined ice lattice parameters which are too large (relying on the synchrotron data given in [18]) and point to a temperature of 135 ± 10 K at the sample, possibly depending on the location of the sample in the $N_2$ stream.

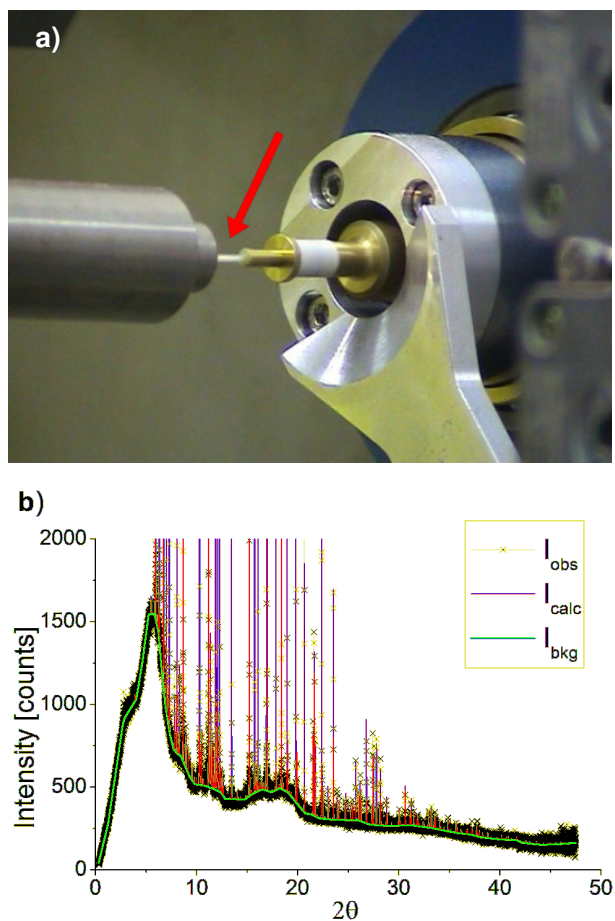

Figure 1. a) Sample stage at ID31/ ESRF (Grenoble): spinning quartz glass capillary (red arrow) mounted in a brass holder partially inserted into the exit tube of a cryo-jet; b) Manually adjusted background (shown as green line) of an ice-containing $CO_2$-$D_2O$-hydrate sample synthesized at 185 K and 266 mbar.

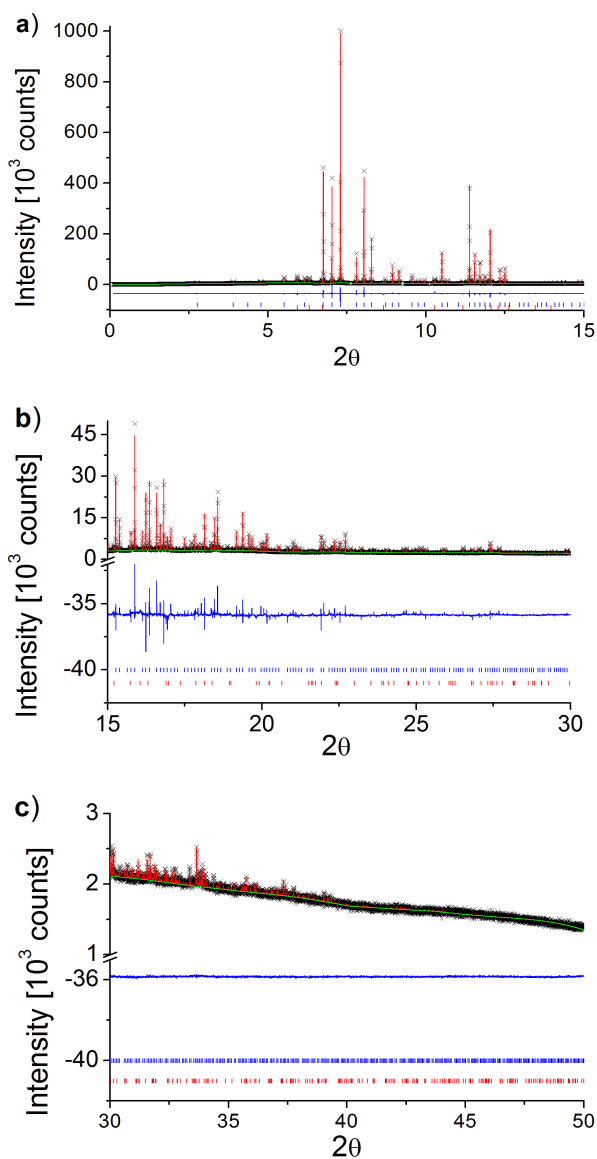

Figure 2. Rietveld-Plot of one $CO_2$-$D_2O$-hydrate synthesized at 30 bars divided into three parts a,b and c for the different $2\theta$ ranges (crosses – observed intensities; red line – calculated intensities; blue line – difference between observed and calclated intensities; tick marks in blue for $CO_2$-hydrate and in red for ice $I_h$).

**Refinements**

All data were analyzed in full-pattern crystallographic structure refinements using the program GSAS [19]. Zero shift, lattice parameters and angle-dependent profile functions with Lorentzian and Gaussian components were refined. The background could not properly be fitted by the implemented functions because of the highly irregular diffuse scattering by the glass capillaries. Hence background points had to be set manually and were linearly interpolated in GSAS (Figure 1b). Structural parameters could be determined with high precision because of a very high reflection to parameter ratio and the very high resolution in $2\theta$ (nominal instrumental contribution to peak broadening is 0.003°$2\theta$; see also Figure 2). Isotropic and in some cases even anisotropic atomic displacement factors ($U_{iso}$, $U_{aniso}$) of guest molecules and cage fillings could be refined simultaneously, which was possible due to the large useful $2\theta$-range of 0-50° (sin $\theta/\lambda$: 1.05 $\text{Å}^{-1}$) with 1472 unique observed hydrate reflections (Figure 2).

**Host model.** Starting structure parameters for the framework were taken from neutron data of $CH_4$-$D_2O$-hydrate synthesized at 60 bars and 273 K [20]. Oxygen framework positions and their $U_{iso}$'s were refined first and finally new framework hydrogen positions were calculated. For this purpose, the refined oxygen and hydrogen positions of $CO_2$-$D_2O$-hydrate synthesized at 30 bars and 273 K and analyzed by neutron diffraction were used [20]; as is customary for X-ray data to account for the density maximum of H-bonded H atoms, the neutron positions were taken and shortened to O-H/D-distances of 0.7 Å.

**Guest model.** The initial guest models for the large and the small cages were also taken from [20]. In the SC the axis of the linear $CO_2$-molecule was set parallel to the [100]-direction pointing to the middle of O1-O1 vectors (Figure 3a), whereby in the LC the C-atom is positioned slightly off-center with surrounding O-atoms (Figure 4a, [20, 3]).

In general this model gave a good fit, but looking closer at the low $2\theta$-angle reflections, significant differences between observed and calculated patterns can be seen (Figure 5a). Therefore other guest models were tried additionally, both for the small cage and the large cage.

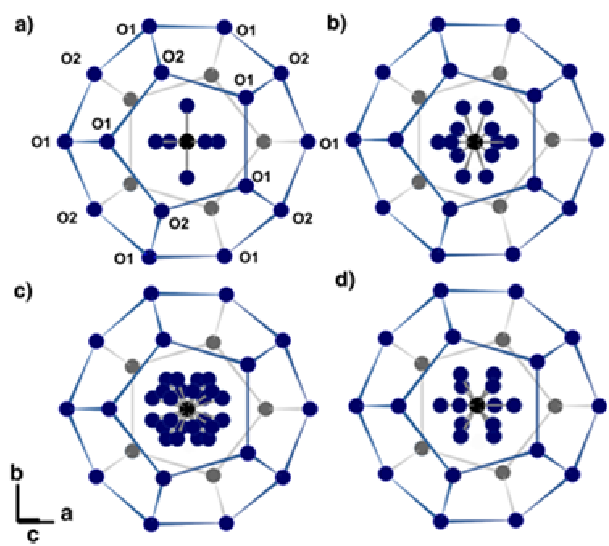

Figure 3. The different models for the $CO_2$-molecule in the small cage drawn with the program POWDER CELL [22] (O-atoms in dark blue in the front [labeling shown in a)] and in grey in the back; C-atoms in black; H-atoms are not displayed): a) $CO_2$-axis pointing to O1-O1-atoms (Multiplicity $M(O_{SC})$ =12; [20]), b) $CO_2$-axis pointing to O1-atoms ($M(O_{SC})$ = 24; [6, 7]), c) $CO_2$-axis pointing to the middle of O1-O2-bonds ($M(O_{SC})$ = 48) and d) $CO_2$-axis pointing to the pentagon faces ($M(O_{SC})$ = 24).

In consideration of chemical and steric arguments the following models are most likely for the SC: the $CO_2$-molecule axis can point either to the O1-atoms (Figure 3b; after Ikeda et al. [6, 7]), to the edges of the small cage defined by O1-O2 vectors (Figure 3c) or to the middle of the pentagonal faces (Figure 3d). In all these models the carbon atom is positioned in the middle of the cage and only the oxygen positions differ. These four models have been tested on $CO_2$-$D_2O$-hydrate synthesized at 30 bars and 271 K and the best fit was achieved when the $CO_2$-molecule axis point either to the O1-O2-cage edges or to the middle of the pentagonal faces. Applying the significance test on the R-values after Hamilton [21] the models after [20] and [6, 7] can be rejected on the 10%-significance level, but it is not possible to decide which one of the other two favored models is the best. However, following the rule of choosing the simpler model for equal fitting, the model with O-atoms pointing to the pentagonal faces is adopted as the best approach for describing the disordered $CO_2$-molecule in the SC at our measuring conditions.

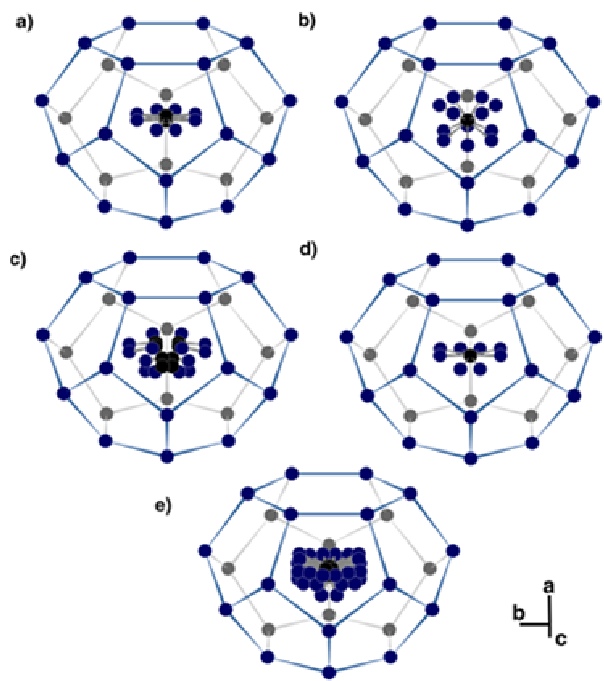

Figure 4. The different models for the $CO_2$-moelucule in the large cage drawn with the program POWDER CELL [22] (O-atoms in blue in the front and grey in the back; C-atoms in black; H-atoms are not displayed): a) C-atom nearly positioned in the cage-centre with $M(C_{LC})$ = 12 surrounded by O-positions in a slightly puckered plane with $M(C_{LC})$ = 48 [20]; b) C-atom positioned in the cage-centre with $M(C_{LC})$ = 6 surrounded by a cloud of O-atoms with $M(O_{LC})$ = 48 [6, 7]; c) C-positions off-centre in two different heights with $M(C_{LC})$ = 48, O-positions in two parallel planes around the C-positions with $M(O_{LC})$ = 48 [11]; d) C-position in the cage-centre with $M(C_{LC})$ = 6, O-positions tilted slightly off-plane in pairs with $M(O_{LC})$ = 24 [8] and e) two C-positions off-centre with $M(C_{LC})$ = 48 surrounded by four different oxygen-positions with $M(O_{LC})$ = 48; [9]).

For the LC, besides the model of [20], the models of Ikeda et al. [6, 7], Takeya et al. [11], Henning et al. [8] and Udachin et al. [9], where position and hence the multiplicity of C- and O-atoms differ considerably (Figure 4), are tested on the same $CO_2$-$D_2O$ hydrate sample synthesized at 30 bars and 271 K. Since there are three times more LC than SC per unit cell the different LC models show a more pronounced influence on the reflections at low 2θ-angles (Figure 5). From a comparison of the recalculated and observed patterns with each other, one can clearly see that the LC models after [20] and [8] fit best. The other models can be rejected at the 10%-significance level applying the significance test for R-values [21]. Yet, while the O-position after Henning et al. [8] in the LC could be refined freely resulting in a reasonable C-O-distance of 1.18(5) for $CO_2$-$D_2O$ hydrate, our starting model [20] does not allow this free refinement. Hence, the $CO_2$-model for the large cage after Henning et al. [8] is preferred, indeed, in the last step of the refinement the displacement parameter of the oxygen in the LC could be refined even anisotropically.

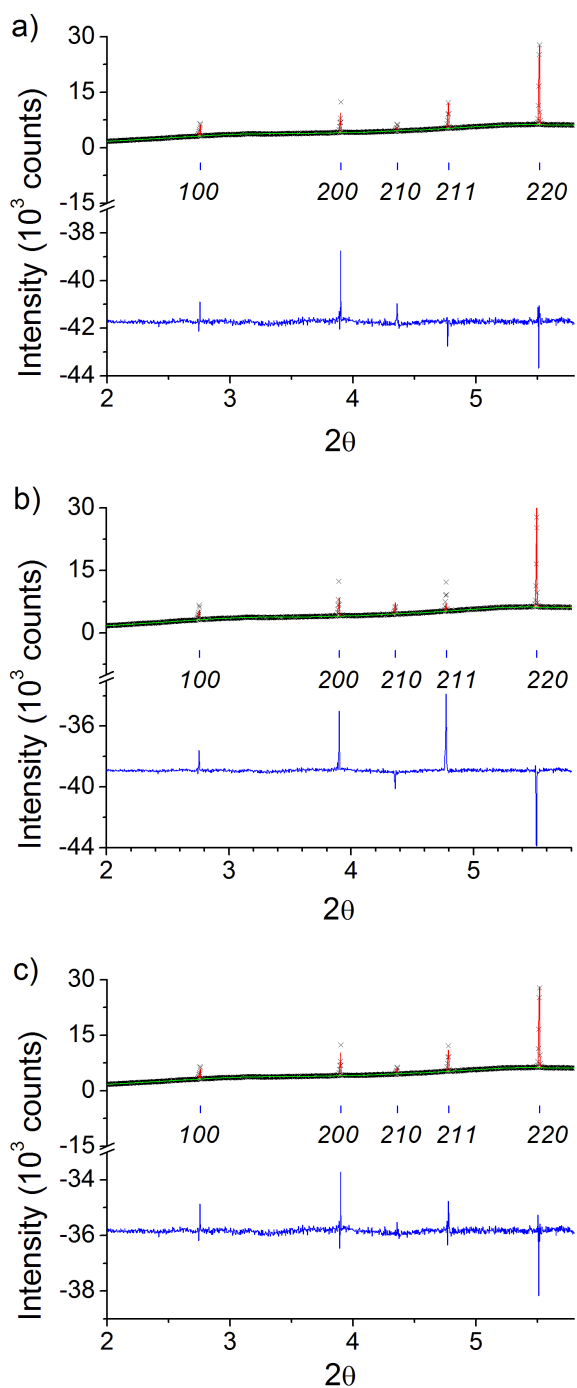

Figure 5. Examples of the fitted low 2θ-angle reflections: a) with $CO_2$-molecules in the small cage after this study, in the large cage after [20]; b) small cage after this study, large cage after [9] c) small cage model after this study, large cage after [8] with refined $CO_2$ oxygen position. Please note the different scales in the difference plots.

The $CO_2$-$D_2O$-hydrate samples formed at temperatures of 185 – 195 K contained such a high amount of ice, that it was not possible to refine the cage fillings of both LC and SC freely. Therefore the filling of the LC was fixed to 100%. Fortunately, one data set of $CO_2$-hydrate sample formed at 195 K shows sufficient quality to refine the occupancies in the SC, the position and $U_{iso}$'s of the framework oxygen-atoms and the $U_{aniso}$ parameters for the oxygen atom in the LC. These refined structural parameters were then taken for the other low temperature samples, allowing refinement of their SC fillings.

**Ice Ih.** The initial structural parameters were taken from Goto et al. [23]. Lattice parameters, $U_{iso}$ and positional parameters of the oxygen atom could be refined in all cases.

**Solidified $CO_2$-gas.** A small amount of 1 – 2 wt.% of solid $CO_2$ was registered in two samples (Table 1), whereas only the lattice parameters and profile parameters of the solidified $CO_2$-gas was refined.

**RESULTS AND DISCUSSION**
Detailed structural parameters are given for one $CO_2$-$H_2O$- and one $CO_2$-$D_2O$-hydrate in Table(s) 2 and 3, respectively. The cage occupancies of all examined $CO_2$-hydrates are listed in Table 1. It should be noted that the low-temperature data taken at 185 K and 190 K suffer from an incomplete reaction and contain only some 15-20% hydrate together with a large amount of ice Ih. Correspondingly, the information on the hydrate structure is less reliable. The esd's of the oxygen positional parameters are smaller than for any other published work, improving by a factor of 2 to 3 over single crystal work [9]. The water framework structure is thus determined with high precision and worth to be analyzed in further detail. It is remarkable that the H-bonded oxygen-

| Frame-work | $T_S$ [K] | $P_S$ [bar] | $f_S$ [bar] | $R_{wp}$ | $a$ | $\theta_{SC}$ [%] | $\theta_{LC}$ [%] | $n$ | $m_{ice}$ [wt.%] | $m_{CO2(s)}$ [wt.%] | $n_{scan}$ |
|---|---|---|---|---|---|---|---|---|---|---|---|
| $H_2O$ | 268.15 | 15 | 13.2[a] | 7.11 | 11.84823(2) | 59.4(5) | 99.2(2) | 6.44(4) | 10.72(5) | - | 9 |
| $H_2O$ | 268.15 | 30 | 23.0[a] | 7.10 | 11.85178(2) | 68.5(4) | 98.4(3) | 6.32(3) | 15.59(4) | 1.04(3) | 16 |
| $D_2O$ | 271.15 | 30 | 23.9[a] | 4.79 | 11.85197(1) | 67.4(3) | 98.9(1) | 6.32(2) | 4.35(4) | - | 26 |
| $D_2O$ | 185 | 0.266 | 0.262[d] | 9.24 | 11.8607(2) | 44(4) | 100.0(5)[b] | 6.7(4) | 86.36(3) | 2.01(6) | 6 |
| $D_2O$ | 190[c] | 0.36 | 0.357[d] | 10.83 | 11.85627(9) | 74(2) | 100.0(5)[b] | 6.14(9) | 78.59(4) | - | 3 |
| $D_2O$ | 190[c] | 0.36 | 0.357[d] | 13.83 | 11.85627(9) | 74(2) | (100.0(5))[b] | 6.14(9) | 84.39(3) | - | 3 |
| $D_2O$ | 190 | 0.36 | 0.357[d] | 10.51 | 11.8580(1) | 77(2) | 100.0(5)[b] | 6.1(1) | 82.37(3) | - | 9 |
| $D_2O$ | 195 | 0.505 | 0.495[d] | 8.94 | 11.86081(4) | 77.8(4) | 100.0(5)[b] | 6.09(2) | 26.71(6) | - | 12 |

[a] Fugacities are taken from [25]
[b] Fixed at 100%, error estimated
[c] Same sample, but 2 different synchrotron datasets refined together
[d] Fugacities calculated after Falenty et al. [15]

Table 1. Results of the crystal structure analysis of $CO_2$-$H_2O$/$D_2O$ hydrate (all structure type I): lattice parameter $a$, cage fillings $\theta$ and hydration number $n$ in dependence of framework water molecules, synthesizing temperature $T_S$ and pressure $P_S$ or fugacity $f_S$. Also listed are weighted R-value $R_{wp}$, amount of ice $m_{ice}$ and remaining, solidified $CO_2$-gas $m_{CO2(s)}$ as well as number of scans $n_{scan}$ (esd's are quoted in parentheses).

oxygen distances vary only very little around 2.75-2.76 Å (Table 4). The H-bonded distances in $CO_2$-hydrate are thus very similar to the ones in ice Ih [24]. The main difference between this structure I hydrate lattice and ice Ih thus resides in the H-bonded O-O-O angles, showing a much larger spread in the clathrate case. The isotopic difference between the oxygen positions in hydrogenated and deuterated material are very small and point to somewhat longer distances in the deuterated compound. Various models for the positioning of the guest molecules were tested; the high quality of the intensity data allowed for a discrimination in terms of the quality of profile fit ($R_{wp}$-values). The disorder of the guest molecules certainly is complex and, at least for the large cages, it is likely that the center of gravity is not in the cage center [11] as in the case of $N_2$-hydrate [26]. It should be noted, that the model deduced from the direct space approach in [11] has arbitrarily fixed ADP's with values that are unphysically low (only 5% of the values for the water oxygen atoms, while one can expect for the guest ADP's values similar to the ADP's of the water framework oxygens). An appropriate disorder model with adequate atomic displacement parameters (ADP's) attached to each atomic position is expected to give an accurate total electron density distribution within the cage. The total weight of each of these densities can then be freely refined via the occupancy factor. Indeed, the free refinement of ADP's given as $U_{iso}$ (see Table 2 and 3) for the guest molecules together with the cage occupancies (Table 1) was possible for the first time; the results are physically meaningful with ADP's of the SC $CO_2$-molecules slightly higher than those of the framework water molecules, and about three times higher ADP's of the LC $CO_2$-molecules. Moreover, some of the positional parameters of the disorder model were refined simultaneously with ADP's and occupancies. The 1σ errors of the cage occupancies are in the order of 0.2% for the large cage and 0.4% for the small cage (see Table 1). It should be mentioned that our result is compatible with the cage occupancies reported in [11] amounting to 0.69 for SC and 0.99 for LC, albeit determined with distinctly higher precision in our case. Unfortunately the synthesis conditions were not reported in [11] so that a more detailed comparison cannot be done. The single crystal study on a deuterated crystal [9], prepared at 3 °C at pressures high enough to bring $CO_2$ in the liquid phase, gave cage occupancies of 0.71(3) for SC

| | M | x | y | z | $U_{iso}$ [Å$^2$] |
|---|---|---|---|---|---|
| O1 | 24 | 0 | 0.30944(5) | 0.11646(5) | 0.0296(2) |
| O2 | 16 | 0.18335(3) | 0.18335(3) | 0.18335(3) | 0.0337(2) |
| O3 | 6 | 0.25 | 0 | 0.5 | 0.0301(4) |
| D1 | 16 | 0.2176 | 0.2176 | 0.2176 | 0.03 |
| D2 | 24 | 0.2163 | 0 | 0.4515 | 0.03 |
| D3 | 24 | 0 | 0.3598 | 0.1471 | 0.03 |
| D4 | 24 | 0 | 0.3146 | 0.0574 | 0.03 |
| D5 | 48 | 0.0476 | 0.2778 | 0.131 | 0.03 |
| D6 | 48 | 0.1364 | 0.2161 | 0.1689 | 0.03 |
| C1 | 6 | 0.25 | 0.5 | 0 | 0.101(2) |
| O4 | 48 | 0.2415(6) | 0.4678(6) | -0.0935(3) | 0.09113* |
| C2 | 2 | 0 | 0 | 0 | 0.036(2) |
| O5 | 24 | 0 | 0.0836 | 0.0513 | 0.064(4) |

* $U(O4)$ was calculated anisotropic [in Å$^2$]: $U_{11}$ = 0.108(2), $U_{12}$ = 0.000(5), $U_{13}$ = 0.005(3), $U_{22}$ = 0.94(11), $U_{23}$ = -0.032(5), $U_{33}$ = 0.071(3).

Table 2. Positional parameters of CO$_2$-D$_2$O-hydrate synthesized at 271 K and 30 bars.

| | M | x | y | z | $U_{iso}$ [Å] |
|---|---|---|---|---|---|
| O1 | 24 | 0 | 0.30951(8) | 0.11643(8) | 0.0335(4) |
| O2 | 16 | 0.18317(6) | 0.18317(6) | 0.18317(6) | 0.0390(4) |
| O3 | 6 | 0.25 | 0 | 0.5 | 0.0338(8) |
| H1 | 16 | 0.2176 | 0.2176 | 0.2176 | 0.03 |
| H2 | 24 | 0.2163 | 0 | 0.4515 | 0.03 |
| H3 | 24 | 0 | 0.3598 | 0.1471 | 0.03 |
| H4 | 24 | 0 | 0.3146 | 0.0574 | 0.03 |
| H5 | 48 | 0.0476 | 0.2778 | 0.131 | 0.03 |
| H6 | 48 | 0.1364 | 0.2161 | 0.1689 | 0.03 |
| C1 | 6 | 0.25 | 0.5 | 0 | 0.102(3) |
| O4 | 48 | 0.243(1) | 0.467(1) | -0.0927(8) | 0.09039* |
| C2 | 2 | 0 | 0 | 0 | 0.034(3) |
| O5 | 24 | 0 | 0.0836 | 0.0513 | 0.048(6) |

* $U(O4)$ was calculated anisotropic [in Å$^2$]: $U_{11}$ = 0.112(4), $U_{12}$ = 0.000(5), $U_{13}$ = 0.005(3), $U_{22}$ = 0.94(11), $U_{23}$ = -0.032(5), $U_{33}$ = 0.071(3).

Table 3. Positional parameters of CO$_2$-H$_2$O-hydrate synthesized at 268 K and 30 bars.

and 1.00(2) for LC, slightly higher than our results. Yet one has to keep in mind that fugacities in this case were somewhat higher, which is expected to result in higher occupancies. Likewise, the value of 0.70 reported in our early work [2] is not too far off from the value of 0.674 for the deuterated sample prepared at identical conditions. We can thus state that the most reliable experimental results so far are in fair agreement. With the unprecedented precision of our latest results one may wish to compare them with thermodynamic predictions. For this purpose we use CSMGem [1] as it allows for the calculation of individual cage occupancies [27] and has proven to be usually better performing than other prediction programs [28]. The results are shown in Fig. 6. The measured and predicted values fall very close, but at least at higher pressure, somewhat outside the experimental error limits. Whether this is due to some inaccuracy of the CSMGem model or due to inaccuracies of the experimental approach cannot be decided easily. Although the long formation times should have brought the reaction rather close to equilibration, reaching thermodynamic equilibrium cannot be fully guaranteed as it needs solid-state diffusion of the constituents through the hydrate lattice, which certainly is not a fast process. One could also turn the argument around and state that the fact that experimental points and predictions are so close is indicative for a situation where experimentally one has reached a point close to equilibrium – giving credit to CSMGem results, due to its remarkable performance in hydrate prediction in general.

The discussion in this paper is limited to the case of CO$_2$-hydrate. Comparable synchrotron data were obtained also for deuterated and hydrogenated $CH_4$-hydrate, deuterated and hydrogenated $N_2$-hydrate and hydrogenated Xe-hydrate synthesized at variable gas fugacities, which will be presented separately in the future.

| Framework | $P_S$ [bar] | $f_S$ [bar] | $T_S$ [K] | d(O1-O1) [Å] | d(O1-O2) [Å] | d(O1-O3) [Å] | d(O2-O2) [Å] |
|---|---|---|---|---|---|---|---|
| H$_2$O | 15 | 13.2 | 268 | 2.759(2) | 2.7521(7) | 2.759(1) | 2.734(2) |
| H$_2$O | 30 | 23.0 | 268 | 2.760(2) | 2.7533(7) | 2.757(1) | 2.744(2) |
| D$_2$O | 30 | 23.9 | 271 | 2.761(1) | 2.7539(4) | 2.7578(6) | 2.737(1) |
| D$_2$O | 0.505 | 0.495 | 195 | 2.773(3) | 2.754(1) | 2.753(2) | 2.735(4) |

Table. 4: O-O-distances of the neighboring framework oxygen atoms for refined CO$_2$-D$_2$O/H$_2$O-hydrate synthesized at different temperatures and pressures.

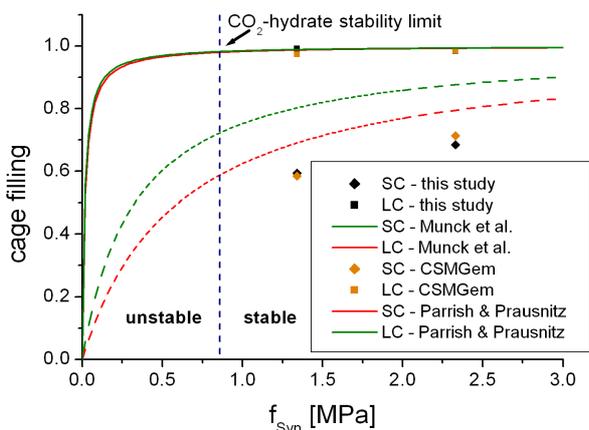

Figure 6. Fugacity dependency of cage filling for $CO_2$-$H_2O$-hydrate. Shown are the experimental cage fillings from this study, Langmuir isotherms after Munck et al. [4] and after Parrish and Prausnitz [5] as well as the calculated cage fillings using CSMGem [1]. The stability limit is calculated also with the program CSMGem [1]. Experimental errors are smaller or equal to the symbol size.

**ACKNOWLEDGEMENTS**
The authors thank the European Synchrotron Radiation Facility (ESRF) in Grenoble for beam-time and support, Heiner Bartels and Eberhard Hensel (Göttingen) for technical support and Michela Brunelli (ESRF) for help and advice on ID31. WFK thanks Keith Hester (at the time at CSM, Golden) for some initial discussion on cage filling issues.